\documentclass[preprint,12pt]{elsarticle}

\usepackage{amssymb}
\usepackage{graphicx}
\usepackage{epstopdf}
\usepackage{lineno}
\usepackage{multirow}
\usepackage{geometry}
\usepackage{subfigure}
\usepackage{booktabs}

\newcommand{\bg}  {counts/(keV$\cdot$kg$\cdot$yr)}
\newcommand{\bbd}    {$0 \nu\beta\beta$}
\newcommand{\qbb}    {$2527.518 \pm 0.013$~keV}

\begin{document}
\begin{frontmatter}

\title{Muon-induced backgrounds in the CUORICINO experiment}



\author[Como,INFNMilano]{E.~Andreotti}
\author[Milano,INFNMilano]{C.~Arnaboldi}
\author[USC]{F.~T.~Avignone~III}
\author[LNGS]{M.~Balata}
\author[USC]{I.~Bandac}
\author[Firenze]{M.~Barucci}
\author[LBNL]{J.~W.~Beeman}
\author[Roma,INFNRoma]{F.~Bellini}
\author[LBNL]{T.~Bloxham}
\author[Milano,INFNMilano]{C.~Brofferio}
\author[LBNL,BerkeleyPhys]{A.~Bryant}
\author[LNGS]{C.~Bucci}
\author[Genova]{L.~Canonica}
\author[Milano,INFNMilano]{S.~Capelli}
\author[INFNMilano]{L.~Carbone}
\author[Milano,INFNMilano]{M.~Carrettoni}
\author[Milano,INFNMilano]{M.~Clemenza}
\author[INFNMilano]{O.~Cremonesi}
\author[USC]{R.~J.~Creswick}
\author[Genova]{S.~Di~Domizio}
\author[LLNL,BerkeleyPhys]{M.~J.~Dolinski}
\author[Wisc]{L.~Ejzak}
\author[Roma,INFNRoma]{R.~Faccini}
\author[USC]{H.~A.~Farach}
\author[Milano,INFNMilano]{E.~Ferri}
\author[Roma,INFNRoma]{F.~Ferroni}
\author[Milano,INFNMilano]{E.~Fiorini}
\author[Como,INFNMilano]{L.~Foggetta}
\author[INFNMilano]{A.~Giachero}
\author[Milano,INFNMilano]{L.~Gironi}
\author[Como,INFNMilano]{A.~Giuliani}
\author[LNGS]{P.~Gorla}
\author[Genova]{E.~Guardincerri}
\author[CalPoly]{T.~D.~Gutierrez}
\author[LBNL,BerkeleyMat]{E.~E.~Haller}
\author[LBNL]{R.~Kadel}
\author[LLNL]{K.~Kazkaz}
\author[Milano,INFNMilano]{S.~Kraft}
\author[LBNL,BerkeleyPhys]{L.~Kogler \corref{contact}}
\author[LBNL,BerkeleyPhys]{Yu.~G.~Kolomensky}
\author[Milano,INFNMilano]{C.~Maiano}
\author[Wisc]{R.~H.~Maruyama}
\author[USC]{C.~Martinez}
\author[INFNMilano]{M.~Martinez}
\author[USC]{L.~Mizouni}
\author[INFNRoma]{S.~Morganti}
\author[LNGS]{S.~Nisi}
\author[Como,INFNMilano]{C.~Nones}
\author[LLNL,BerkeleyNuc]{E.~B.~Norman}
\author[Milano,INFNMilano]{A.~Nucciotti}
\author[Roma,INFNRoma]{F.~Orio}
\author[Genova]{M.~Pallavicini}
\author[Legnaro]{V.~Palmieri}
\author[Milano,INFNMilano]{L.~Pattavina}
\author[Milano,INFNMilano]{M.~Pavan}
\author[LLNL]{M.~Pedretti}
\author[INFNMilano]{G.~Pessina}
\author[INFNMilano]{S.~Pirro}
\author[INFNMilano]{E.~Previtali}
\author[Firenze]{L.~Risegari}
\author[USC]{C.~Rosenfeld}
\author[Como,INFNMilano]{C.~Rusconi}
\author[Como,INFNMilano]{C.~Salvioni}
\author[Wisc]{S.~Sangiorgio}
\author[Milano,INFNMilano]{D.~Schaeffer}
\author[LLNL]{N.~D.~Scielzo}
\author[Milano,INFNMilano]{M.~Sisti}
\author[LBNL]{A.~R.~Smith}
\author[LNGS]{C.~Tomei}
\author[Firenze]{G.~Ventura}
\author[Roma,INFNRoma]{M.~Vignati}

\address[Como]{Dip.\ di Fisica e Matematica dell'Univ.\ dell'Insubria, Como I-22100 - Italy}
\address[INFNMilano]{Sez.\ INFN di Milano Bicocca, Milano I-20126 - Italy}
\address[Milano]{Dip.\ di Fisica dell'Univ. di Milano-Bicocca I-20126 - Italy}
\address[USC]{Dept.\ of Physics and Astronomy, Univ.\ of South Carolina, Columbia, SC 29208 - USA}
\address[LNGS]{Laboratori Nazionali del Gran Sasso, Assergi (L'Aquila) I-67010 - Italy}
\address[Firenze]{Dip.\ di Fisica dell'Univ. di Firenze and Sez.\ INFN di Firenze, Firenze I-50125 - Italy}
\address[LBNL]{Lawrence Berkeley National Lab., Berkeley, CA 94720 - USA}
\address[Roma]{Dip.\ di Fisica dell'Univ. di Roma La Sapienza, Roma  I-00185 - Italy}
\address[INFNRoma]{Sez.\ INFN di Roma, Roma  I-00185 - Italy}
\address[BerkeleyPhys]{Dept.\ of Physics, Univ.\ of California, Berkeley, CA 94720 - USA}
\address[Genova]{Dip.\ di Fisica dell'Univ. di Genova and Sez. INFN di Genova, Genova I-16146 - Italy}
\address[LLNL]{Lawrence Livermore National Laboratory, Livermore, CA, 94550 - USA}
\address[Wisc]{Univ.\ of Wisconsin, Madison, WI 53706 - USA}
\address[CalPoly]{California Polytechnic State Univ., San Luis Obispo, CA 93407 - USA}
\address[BerkeleyMat]{Dept.\ of Materials Science\ and Enginineering, Univ.\ of California, Berkeley, CA 94720 - USA}
\address[BerkeleyNuc]{Dept.\ of Nuclear Engineering, Univ.\ of California, Berkeley, CA 94720 - USA}
\address[Legnaro]{Laboratori Nazionali di Legnaro, Legnaro (Padova) I-35020 - Italy} 

\cortext[contact]{Corresponding author. Tel. 510.486.4034, e-mail lkogler@berkeley.edu}

\begin{abstract}
To better understand the contribution of cosmic ray muons to the CUORICINO background, ten plastic scintillator detectors were installed at the CUORICINO site and operated during the final 3 months of the experiment.  From these measurements, an upper limit of 0.0021 \bg ~(95\% C.L.) was obtained on the cosmic ray induced background in the neutrinoless double beta decay region of interest.  The measurements were also compared to \textsc{Geant4} simulations.
\end{abstract}

\end{frontmatter}

\begin{linenumbers}
\section{Introduction}
\label{sec:intro}

Understanding the nature of neutrino mass is one of the key topics at the frontier of fundamental physics. One of the best opportunities for investigating this problem is searching for neutrinoless double beta decay (\bbd), a transition in which a nucleus (A, Z) decays into a daughter (A, ${\textnormal Z}+2$) with the emission of two electrons but no (anti-)neutrinos.

The CUORICINO experiment was a $^{130}$Te-based search for \bbd.  It consisted of an array of 62 tellurium dioxide (TeO$_2$) bolometers with a total mass of 40.7 kg. It was operated at the Laboratori Nazionali del Gran Sasso (LNGS) in Assergi, Italy, from early 2003 to June 2008.  The CUORICINO detector was built as a prototype for the CUORE experiment, which will have 19 CUORICINO-like towers and is presently under construction at LNGS.

The CUORICINO crystals were arranged in a tower made of 13 levels, 11 with four 5$\times$5$\times$5 cm$^3$ crystals and 2 with nine 3$\times$3$\times$6 cm$^3$ crystals.
Each crystal was operated as a bolometer able to detect an energy deposition by recording the resulting temperature increase with a neutron transmutation doped Ge thermistor \cite{Arnaboldi:2008}.  In the case of \bbd, the summed  energies of the  electrons and recoiling nucleus would result in a  mono-energetic peak at the \bbd ~transition energy of \qbb ~for $^{130}$Te \cite{qvalue}.

\begin{figure}[hbt]
   		\begin{center}
        \ifx\pdfoutput\undefined 
        \else
        \includegraphics[width=0.90\textwidth]{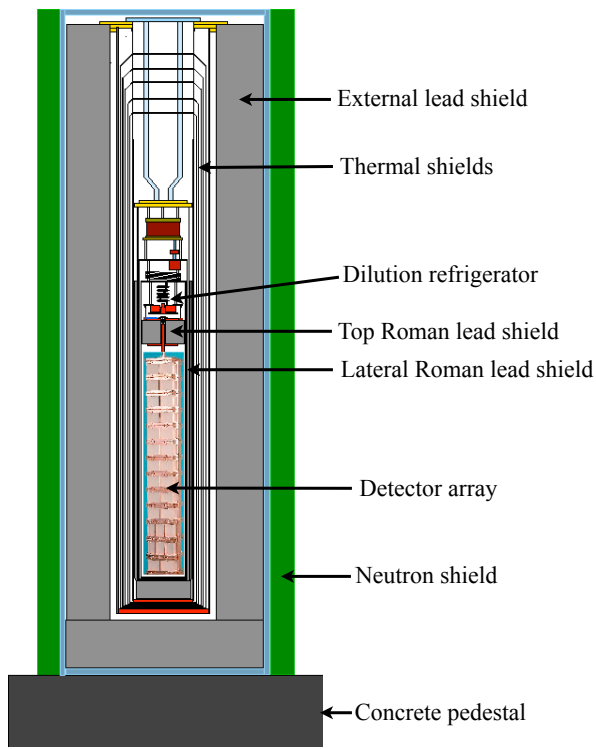}
        \fi
    \end{center}
    \caption{The layout of CUORICINO showing the tower, the various heat shields, and the external shielding}
    \label{fig:cuoricino}
\end{figure}

The detector operated at $\sim$10~mK, cooled by a dilution refrigerator and surrounded by several layers of shielding.  Directly above the detector was a 10 cm thick layer of low-activity ``Roman" lead (from ancient Roman shipwrecks).  Around the sides of the detector were several layers of thermal shields and a 1.2 cm thick cylindrical Roman lead shield. The thermal shields were made from electrolytic copper and totaled at least 1.5 cm in thickness.  
Outside the cryostat was a 10~cm low-activity lead shield and a 10~cm standard lead shield.  The cryostat and shields were surrounded by a Plexiglas box flushed with clean N$_{2}$ from a liquid nitrogen evaporator to avoid radon, followed by a 10 cm borated polyethylene neutron shield.  
A top lead shield was located about 50 cm above the top plate of the cryostat.
The entire setup was enclosed in a Faraday cage to reduce electromagnetic interference. The assembly is shown in Figure \ref{fig:cuoricino}.
A more detailed description of the detector can be found in Ref.~\cite{Arnaboldi:2008}.

In CUORICINO, any single-bolometer energy deposition in the \bbd ~energy region is a potential background that can decrease the sensitivity of the experiment. Cosmic rays are one source of background.  The 3200~mwe overburden at Gran Sasso eliminates the soft cosmic ray component and reduces the flux of penetrating muons by six orders of magnitude to $\sim$1.1~$\mu/($h$\cdot $m$^2)$ \cite{MACRO1}, with a mean energy of $\sim$270~GeV and an average zenith angle $\langle \theta \rangle\sim35$~degrees. The azimuthal distribution reflects the mountain profile \cite{MACRO2, MACRO3}.

A muon could produce a bolometer signal by interacting directly in the detector.  Additionally, muons interacting in the detector, shieldings, or surrounding materials could create secondary products that might mimic a \bbd ~decay.  For example, neutrons produced by cosmic rays are very energetic and thus difficult to block with shields.  Photons emitted in $(n, n' \, \gamma)$ or $(n, \gamma)$ reactions could appear near the \bbd ~energy.
Neutron production increases with the atomic weight of the material; therefore, lead shields can be a strong source of muon-produced neutrons.
However, neutron production is mostly associated with showers, so this background may be effectively identified by coincident events in different bolometers.

Several  Monte Carlo simulations have been carried out on cosmic ray-induced backgrounds  but few direct measurements have been made \cite{Kudryavtsev:2008fi, Araujo:2008ze, hime, wul2, kud, araujo, wang, bergamasco, gor}.  For the present study, an external muon detector was installed to tag muon-induced background events in CUORICINO during its last three months of operation.

Section \ref{sec:exp_set} and Section  \ref{sec:det_ope} give details of the muon detector setup and performances. Section \ref{sec:simulation} is a summary of the Monte Carlo simulations, while Section \ref{sec:dat_ana} describes the data analysis and results.

\section{Muon Detector Setup}
\label{sec:exp_set}

\begin{figure}[hbt]
   		\begin{center}
        \ifx\pdfoutput\undefined 
        \else
        \includegraphics[width=0.90\textwidth]{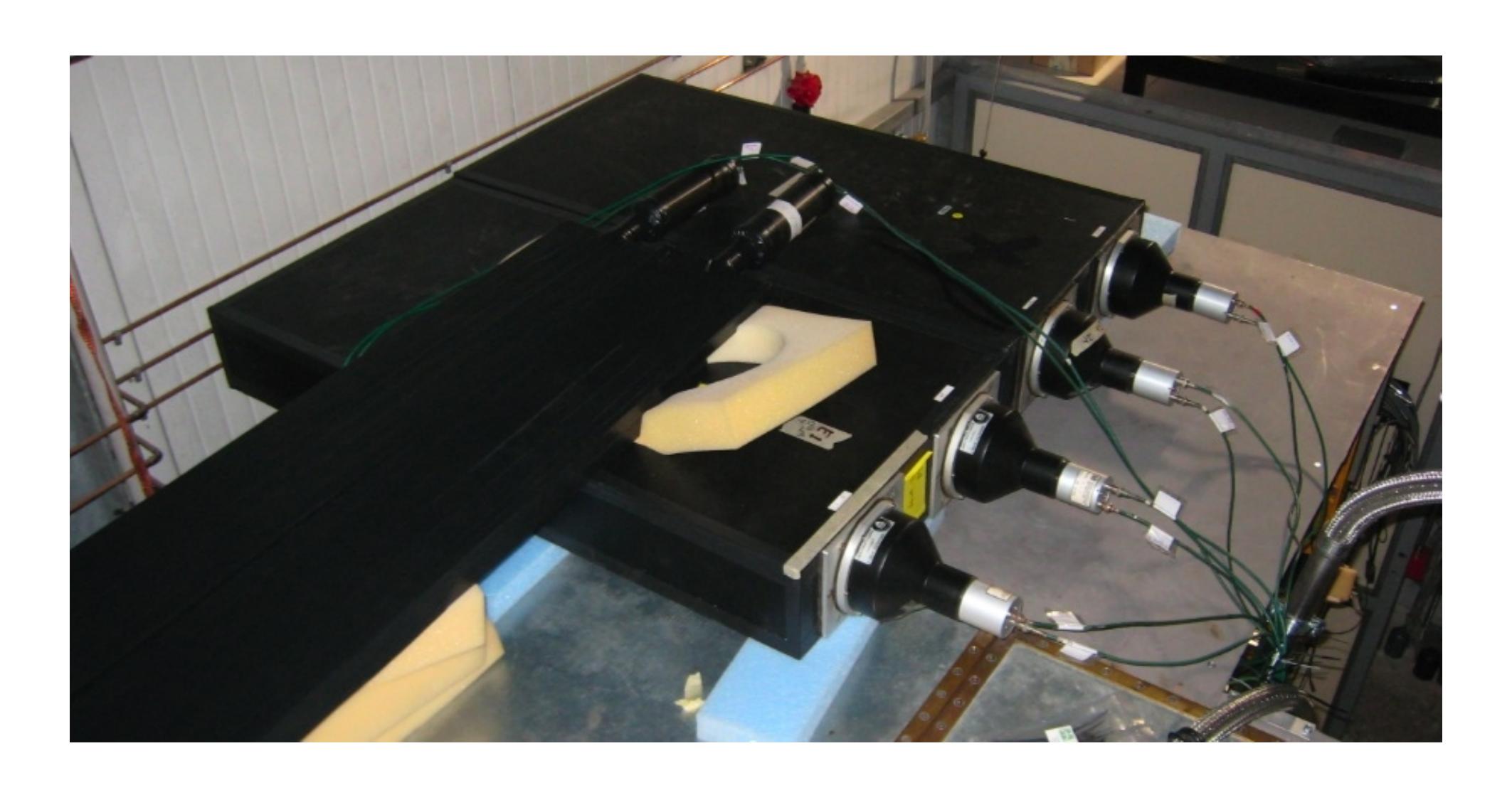}
        \fi
    \end{center}
    \caption{Four of the ten scintillators used (types B and D), shown from the top of the CUORICINO Faraday cage}

    \label{fig:fotoLivermore}
\end{figure}

An array of ten large plastic scintillators placed outside of the Faraday cage, which surrounds the detector, was used to tag muons. The scintillation counters were obtained from previous experiments; the various types are described in Table \ref{ScintillatorsProperties}.  The total sensitive surface area of the scintillators was about 3.67 m$^2$.  A photograph of four of the scintillators is shown in Figure \ref{fig:fotoLivermore}.

\begin{table}[!hbt]
	\small
\label{tab:scintillators}
\centering
\begin{tabular}[c]{c c c c c }
\toprule \addlinespace
Scintillator & Length & Width & Thickness  & Number  \\
Label & (cm) & (cm) & (cm) & of PMTs \\
\midrule
A1 & $100$ & $50$ & $5$ & $1$\\
A2 & $100$ & $50$&$5$ & $1$\\
B1 & $120$ & $60$ & $15$ & $2$\\
B2 & $120$ & $60$ & $15$ & $2$\\
C1 & $96$  & $42.5$ & $3.2$ & $1$\\
C2 & $55$  & $64$  & $3.2$ & $1$\\
D1 & $200$ & $20$ & $3$ & $1$\\
D2 & $200$ & $20$ & $3$ & $1$\\
D3 & $200$ & $20$ & $3$ & $1$\\
D4 & $200$ & $20$ & $3$ & $1$\\
\bottomrule

\end{tabular}
\caption{Dimensions of the plastic scintillators used \label{ScintillatorsProperties}}
\end{table}

\begin{figure}[hbt]
   \begin{center}
	\begin{tabular}[c]{c c}
        \ifx\pdfoutput\undefined 
        \else
        \includegraphics[trim = 6cm 2cm 4cm 1cm, clip, width=0.55\textwidth]{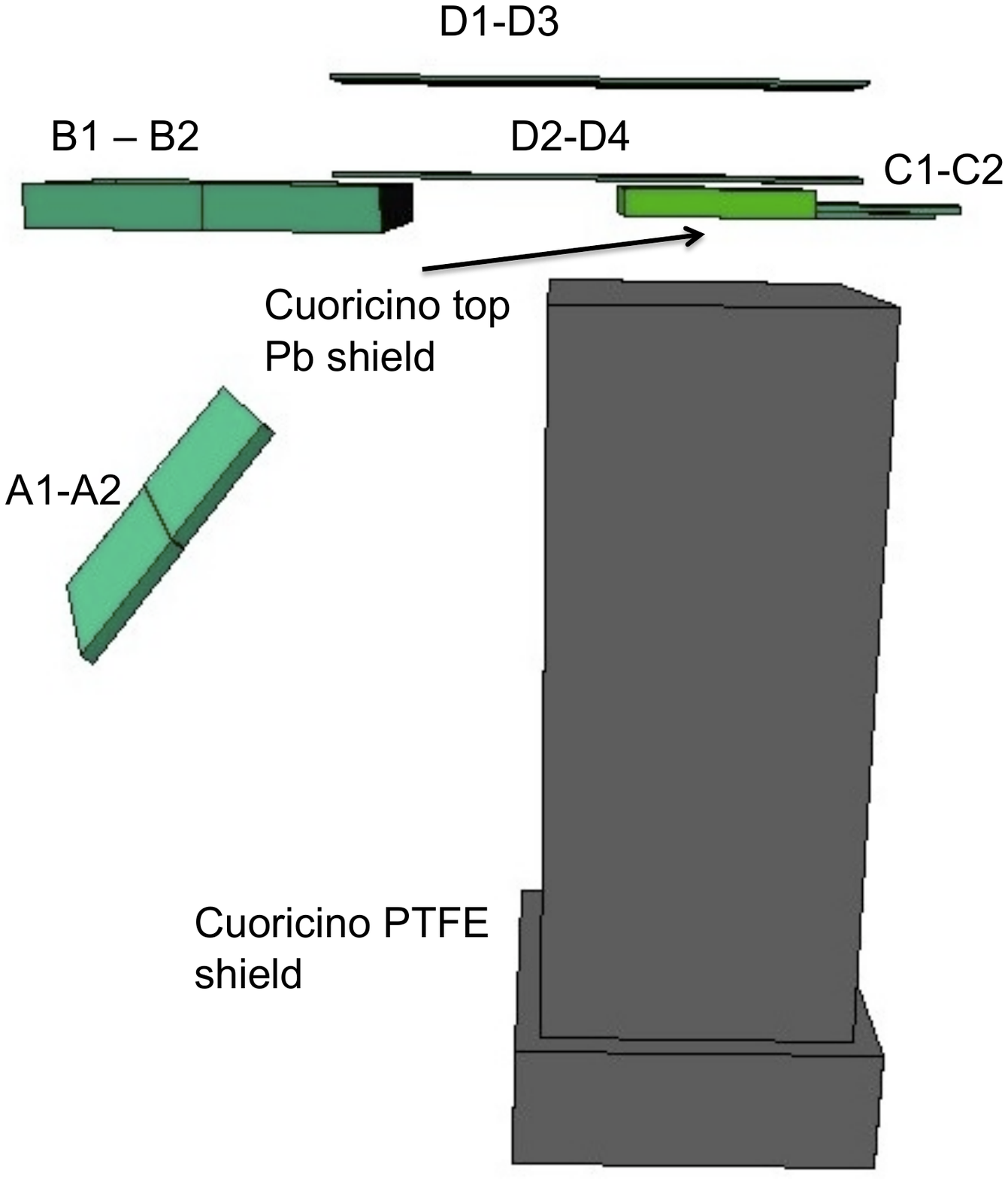}
		\includegraphics[trim = 6cm 2cm 4cm 2cm, clip, width=0.50\textwidth]{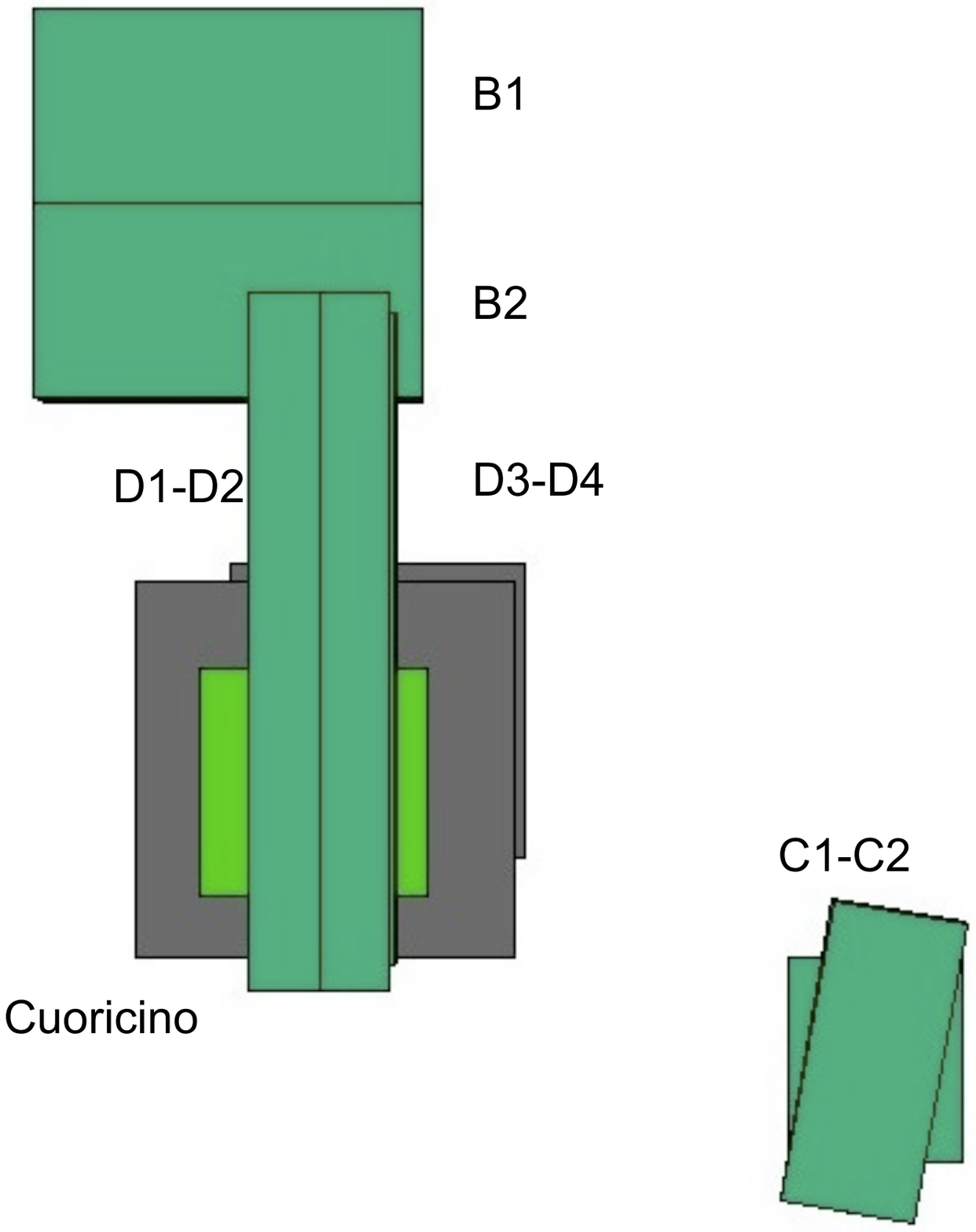}
        \fi
\end{tabular}
    \end{center}
\caption{The drawings show the positions of the scintillators around CUORICINO (left: side view, right: top view): the large, dark grey box is the neutron shield placed around the detector, and the smaller objects are the scintillators and the top part of the lead shield.  The support structures for the scintillators have been omitted.}
	\label{fig:scintPlacem}
\end{figure}

The scintillators were deployed to tag as many as possible of the muons hitting the lead shields while accounting for both the angular distribution of the incoming muons and the geometric constraints from existing structures.
A simple Monte Carlo simulation reproducing the muon flux measured by MACRO \cite{MACRO3} was used to optimize the placement of the counters. 
The arrangement of the scintillators is shown in Figure \ref{fig:scintPlacem}.

\begin{figure}[hbt]
   \begin{center}
        \ifx\pdfoutput\undefined 
        \else
        \includegraphics[width=\textwidth]{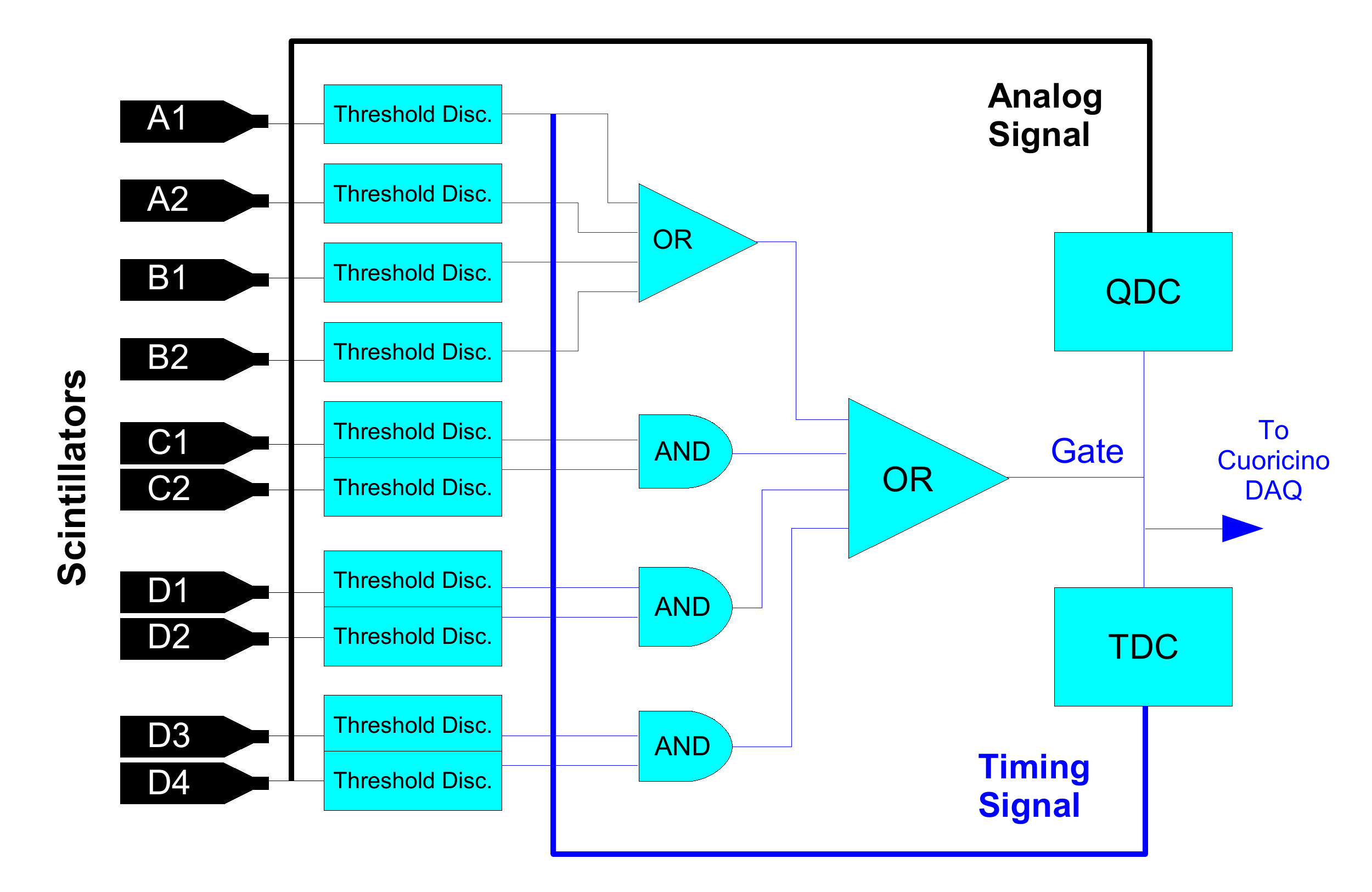}
        \fi  
    \caption{Principle of operation of the electronics and the DAQ system}
    \label{fig:analog_scheme}
    \end{center}
\end{figure}

Each scintillator was read out by one photomultiplier tube (PMT) attached to one of its smallest faces, except the type B scintillators which had two PMTs on the same face with their outputs summed.  

The type A and B scintillators were the thickest and were operated alone.  For these scintillators, the energy released by a through-going minimum ionizing particle was greater than 8 MeV, which was well above any naturally occurring gamma or beta background as well as most naturally occurring alpha lines; therefore, muons may be 
discriminated from background by simply applying cuts on the energy. 
The type C and D scintillators were about 3 cm thick and were operated in pairs.  For each pair, one scintillator was stacked on top of the other and a trigger signal was generated only when they were hit in coincidence (within 120 ns of each other), as indicated in Figure \ref{fig:analog_scheme}.  A 5~cm thick layer of lead was placed between each pair of type D scintillators to further reduce backgrounds.

The signals from the PMTs were sent to the electronics and data acquisition (DAQ) systems.
The analog electronics stage, constructed from commercial NIM modules, was responsible for generating the trigger signals; the analog signals were afterwards digitized by a dedicated VME data acquisition system synchronized with the CUORICINO DAQ (Figure \ref{fig:analog_scheme}).  Each PMT signal was split in two copies: one was sent to a threshold discriminator; the other, after being delayed, was fed into a VME QDC board (Caen V792 N).  The QDC board recorded the charge from the PMT (integrated over 120 ns), which was proportional to the energy released in the scintillator.  The logic signals from the threshold discriminators were also split: one copy went to the NIM boards implementing the trigger logic, while the other went to a VME TDC board (Caen V775 N).  The TDC board recorded the relative time between all PMT signals and the trigger, with a nominal precision of 70 ps.  However, since the typical time resolution of the PMTs was 1--2 ns, the relative time between multiple PMT hits associated with a single trigger is known to a few ns, while the absolute trigger time is known only to the precision of the CUORICINO DAQ (8 ms).  

\section{Detector Operation and Performance}
\label{sec:det_ope}
The muon tagging system was operated with CUORICINO from 12 March to 26 May 2008.  The system was running $\sim$53\% of the time because of CUORICINO calibrations and downtime for repairs and maintenance; the total live time was 38.6 days. 

\begin{figure}[ptb]
  \begin{center}
	\includegraphics[angle=90, width=0.8\textwidth]{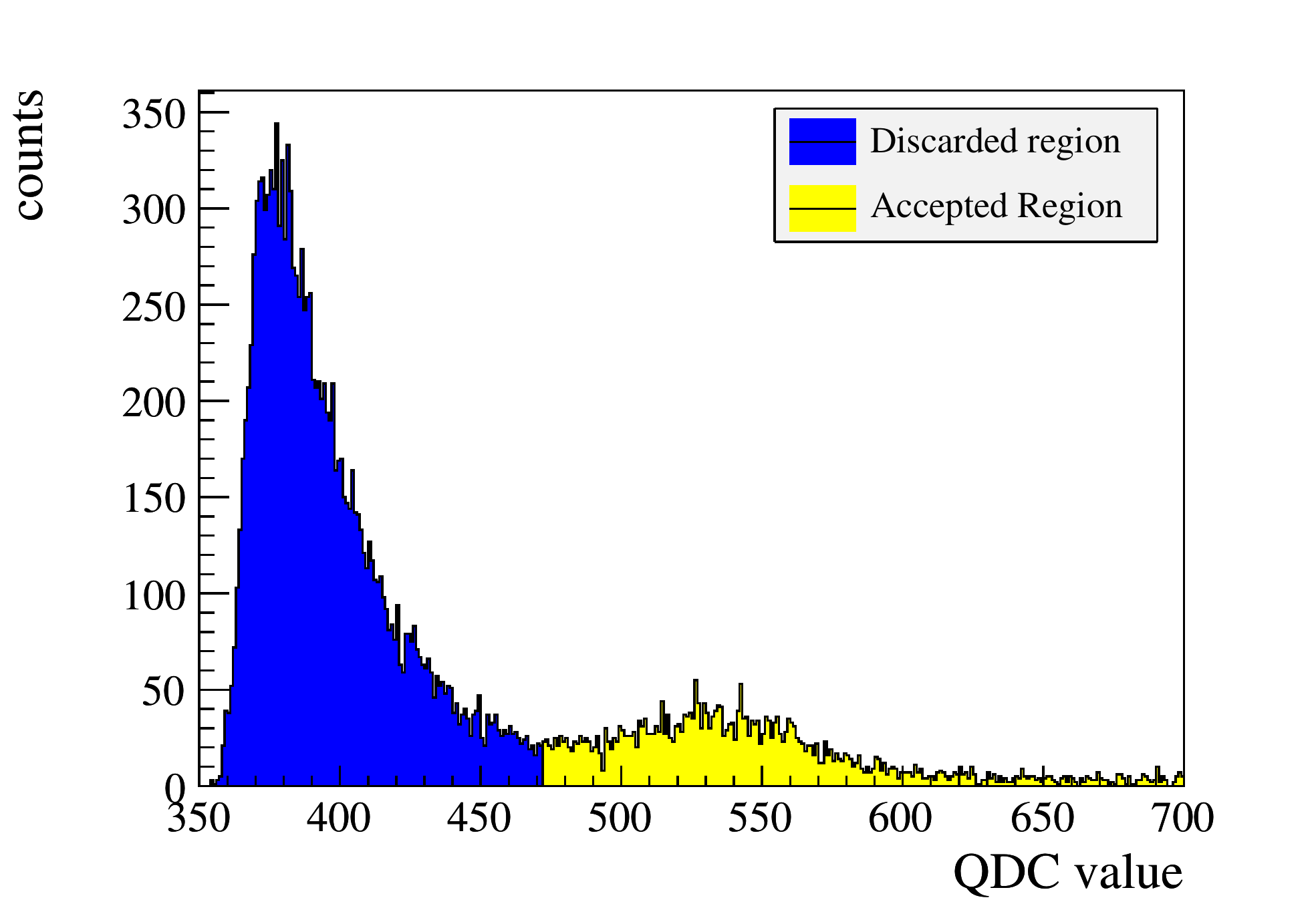}
    \caption{Energy spectrum acquired underground at LNGS using detector A1.  The darker region (not used in the analysis) is dominated by low energy background, while lighter region primarily contains muon events.
 The X axis is QDC counts, proportional to energy.}\label{fig:scint_spectrum}
  \end{center}
\end{figure}

Figure \ref{fig:scint_spectrum} shows the energy spectrum acquired by one of the type A scintillators. Two regions are evident: 
a low energy background region and a broad peak at higher energies.
The low energy background is due to radioactivity, dark noise, and muons that clip the scintillator, whereas the higher energy peak is mostly due to cosmic ray muons.

The efficiencies of the detectors were measured above ground in the assembly hall of LNGS, where the muon rate was much larger.  The measurement was made by placing a pair of scintillators (A and B) above and below the scintillator whose efficiency was being measured (C), such that any muon passing through both A and B must also pass through C.  If $N_{AB}$ is the number of hits occurring in coincidence in detectors A and B, and $N_{ABC}$ is the number of hits in coincidence between all 3 detectors then the efficiency of detector C is simply $\eta_C = N_{ABC}/N_{AB}$. Of course, the efficiency depends on the thresholds set by the threshold discriminators; the thresholds were chosen based on these measurements to be as high as possible while still maintaining an efficiency close to unity.  The individual efficiencies of the detectors measured in this fashion were generally greater than 95\%; however, there was an additional loss of efficiency from cuts applied in the analysis to reduce background (described below).

Muons were discriminated from the background in the thick scintillators (types A and B) with an energy cut: for these detectors a software threshold was set at the dip between the signal and background regions in the spectrum (Figure \ref{fig:scint_spectrum}).  A rough estimate of the loss of efficiency due to this cut was obtained by assuming a Gaussian shape for the muon peak.  
For the type A (5~cm thick) detector shown in Figure \ref{fig:scint_spectrum}, this cut rejected $\sim$10\% of the muons, although for the thicker type B (15~cm) detectors, the estimated loss of efficiency was less than 1\%. 

For the thin scintillators, coincidences between different detectors were used to generate triggers as described in 
Section \ref{sec:exp_set} and no further cut on the energy of the events was applied in the analysis, since the muon peak was not well separated from the background in the energy spectrum.

In order to determine the overall efficiency of the setup for tagging muons associated with CUORICINO bolometer events, the intrinsic detector efficiency and the efficiency of the software cuts were combined for each detector.   This information was then included in a Monte Carlo simulation precisely reproducing the CUORICINO geometry, the positions of the scintillators, and the distribution of the muon flux.  The simulation is described in more detail in Section \ref{sec:simulation}.

The total trigger rate of the muon detectors combined was $\sim$14~mHz with no cuts applied, or $\sim$4~mHz with energy threshold cuts, while the expected signal rate from the simulation was 1.75 mHz.  The difference between the predicted and measured rate is due to the fact that the trigger thresholds were kept low in order to maximize the efficiency of muon detection; this resulted in the inclusion of some triggers caused by radioactive decays and dark noise.  These spurious muon triggers contributed some background to our measurement through the increased rate of accidental coincidences between the scintillators and bolometers, which was taken into account in the analysis described in Section \ref{sec:dat_ana}.

\section{Simulation}
\label{sec:simulation}

\textsc{Geant4} version 9.2\footnote{A known bug affecting the neutron inelastic interactions has been fixed in \textsc{Geant4} 9.2: http://geant4.cern.ch/support/ReleaseNotes4.9.2.html} \cite{geant} was used to simulate the muon-induced backgrounds in CUORICINO.  The LBE (Low Background Experiment) physics list was used. 
The \textsc{Geant4} capability of event-by-event simulation was employed to follow the whole sequence of secondary tracks from the initial interaction to the detector, including the contribution of neutrons generated from muon interactions in the shields.  The complete structure of the scintillators, external shields, internal shields, and detector geometry was implemented according to the model shown in Figure \ref{fig:scintPlacem}.  The propagation of particles through the rock overburden was not simulated, but was accounted for as described below.

An external code simulated the muon energy and angular distribution in the underground laboratory of LNGS. Muons were generated on a 6~m hemisphere in the underground laboratory according to the angular distribution measured by the MACRO experiment \cite{MACRO3}.  The generated muons were then assigned an energy based on the ground-level energy spectrum for that angle, which was approximated as \cite{ref:amsler}:
\begin{equation}
\frac{dN}{dE_{GL} \cdot d\Omega} \propto 
\frac{0.14 \cdot E_{GL}^{-2.7}}{\rm s\: cm^2\: sr\: GeV} 
\left( \frac{1}{1+\alpha\: E_{GL}\: {\rm cos}\,\theta}+ 
\frac{0.054}{1+\beta\: E_{GL}\: {\rm cos}\,\theta } \right),
\end{equation}
where $E_{GL}$ is the energy at ground level, $\alpha = 1.1/115$~GeV, and $\beta = 1.1/850$~GeV.
The ground-level energy was then translated into an underground energy based on the formula \cite{ref:amsler}:
\begin{equation}
E_U = (E_{GL} + \epsilon)\cdot e^{-bX(\theta,\phi)} - \epsilon,
\end{equation}
where $E_U$ is the energy underground, $b = 0.4 \times 10^{-5}\: {\rm cm}^{2}/{\rm g}$, $\epsilon = 540.0$~GeV, and $X(\theta,\phi)$ is the thickness times density of the overburden in the given direction.
The advantage of this method is that it includes the correlation between the direction and energy of the muons underground.  
The range of above-ground energies simulated was chosen for each direction such that the underground energies spanned from 1~GeV to
2~TeV, which corresponds to $\sim$99\% of the underground muon flux.    

The output of the simulation contained the event number, detector number (scintillator number or bolometer number), hit time, and energy released in the detector. This output was used to produce spectra and scatter plots, taking into account the detector response and analysis cuts in order to reproduce the experimental conditions.  A Gaussian smearing of 8 keV (full-width at half maximum) modeled the bolometer resolution. 

The Monte Carlo simulation produced the equivalent of about 3.5 years of data ($\sim 8 \times 10^6$~primary muons).
In addition to statistics, the simulations were subject to systematic uncertainties: uncertainty in the primary muon flux and spectrum (8\%) \cite{MACRO1}, \textsc{Geant4} electromagnetic tracking (5\%), uncertainty in the muon-induced neutron yield (40\%), and neutron propagation and interaction (20\%) \cite{Pandola}.  Analysis of simulation results will be discussed in Section \ref{sec:sim_results}.

\section{Data Analysis}
\label{sec:dat_ana}
The analysis involved searching for correlations between muon triggers and events in the CUORICINO bolometer array.  A coincidence was defined as a muon detector event occurring within $\pm$50 ms of a bolometer event.  This large window, chosen based on the time resolution of the bolometer signals ($\sim$30 ms), was not a limitation due to the low event rates.

\begin{figure}[ptb]
  \begin{center}
    \includegraphics[angle=90, width=0.75\textwidth]{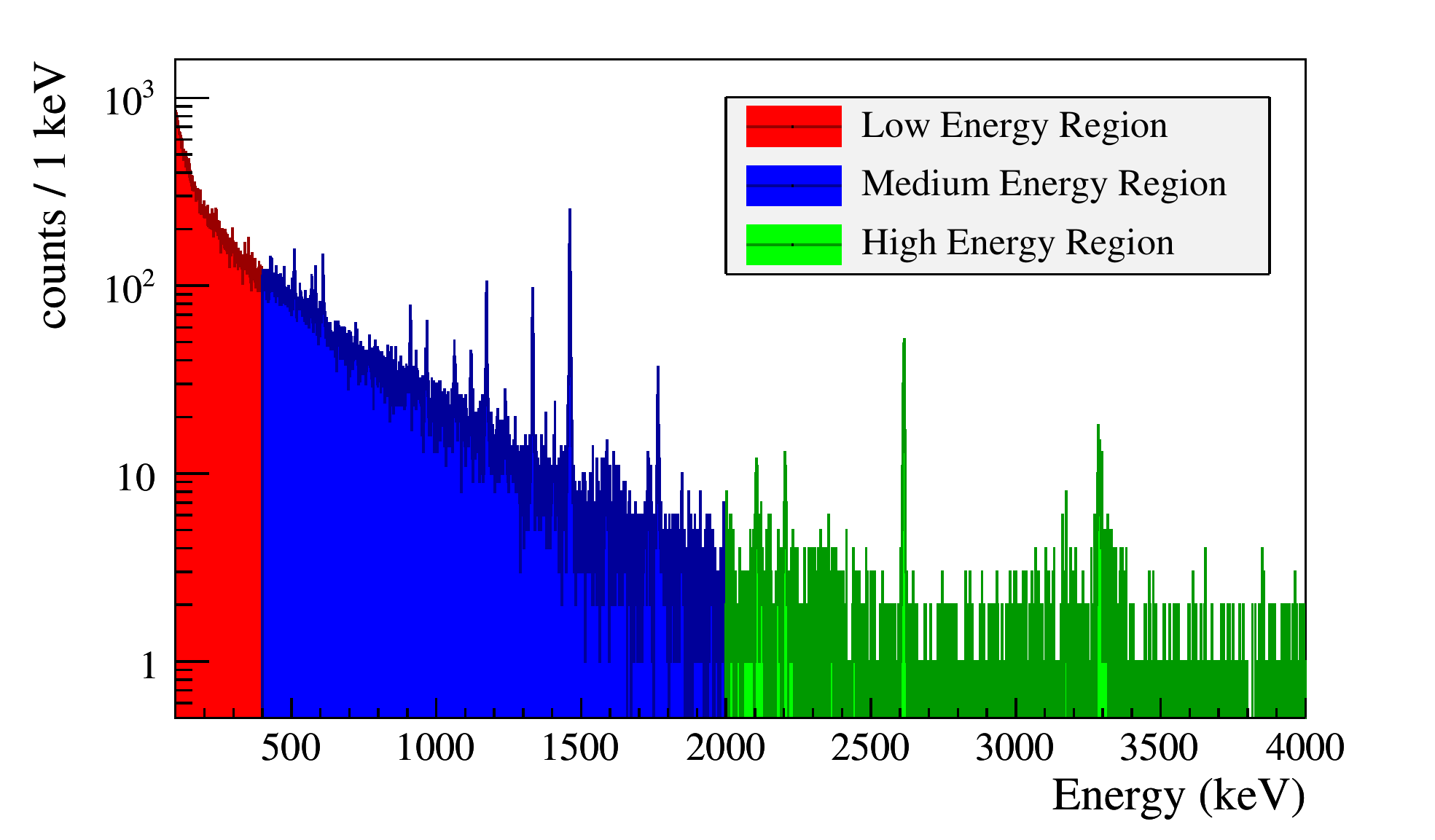}
    \caption{Energy spectrum of the CUORICINO background showing the division of energy regions used in the analysis.  No bolometer anti-coincidence cut has been applied.}\label{fig:spectrum}
  \end{center}
\end{figure}

The bolometer spectrum was divided into three energy regions: 200--400~keV, 400--2000~keV, and 2000--4000~keV, as shown in Figure \ref{fig:spectrum}.  The background rate varies by several orders of magnitude over the complete spectrum; therefore, it is useful to treat the high energy region, which contains the Q-value for \bbd ~decay (\qbb), separately from the lower energy regions where the background is much higher.  In addition to the \bbd ~Q-value, the high energy region contains the $^{208}$Tl $\gamma$ line at 2614.5 keV, the $^{190}$Pt $\alpha$ line at 3249 keV (including nuclear recoil), and an approximately constant background from 3--4 MeV, which is believed to be due to degraded alphas.  This region may also have a cosmic ray component, and was therefore investigated with this measurement.

\begin{figure}[ptb]
  \centering
     \subfigure[Expected number of accidentals vs.~measured coincidences]{
          \label{fig:barchart1}
          \includegraphics[angle=90, width=\textwidth]{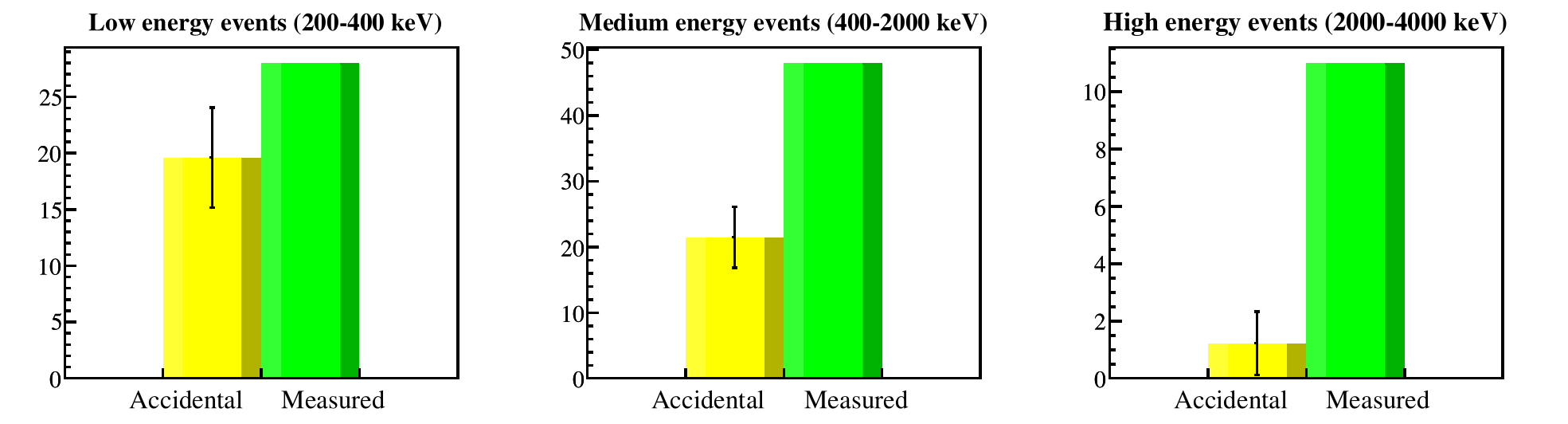}}\\
     \subfigure[Accidentals vs.~measured coincidences with bolometer anti-coincidence cut]{
          \label{fig:barchart2}
          \includegraphics[angle=90, width=\textwidth]{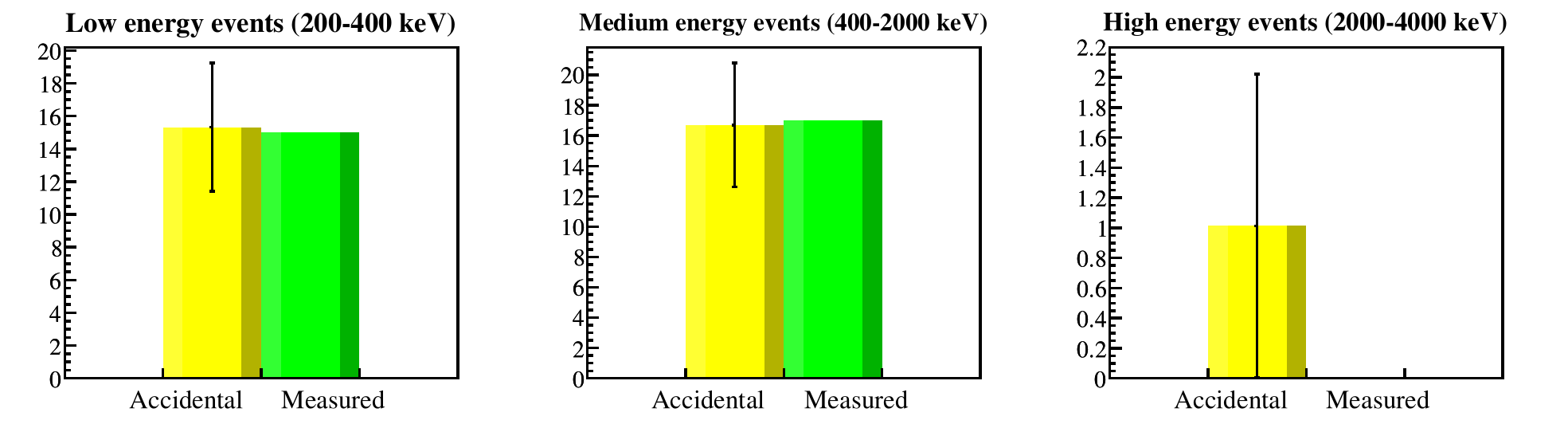}}
     \label{fig:barchartmultifig}
    \caption{Comparison of the expected number of accidental coincidences and the number of observed coincidences between the muon detector and bolometer signals.  The error bar on the accidentals column represents Poisson fluctuations.  Figure \ref{fig:barchart1} includes all bolometer events in the given energy region, while Figure \ref{fig:barchart2} only includes bolometer events which pass an anti-coincidence cut (i.e. they do not occur within 100 ms of any other bolometer event).}  
\end{figure}

In the limit of low rates, the rate of ``accidental'' coincidences between muon events and bolometer events is given by:
\begin{equation}
R_{{\rm accidental}}=2 \cdot R_{{\rm bolo}} \cdot R_{\mu} \cdot \Delta T
\end{equation}
where $R_{{\rm bolo}}$ is the bolometer event rate, $R_{\mu}$=4.01 mHz is the muon rate, and $\Delta T$=50 ms is the width of the coincidence window.
Multiplying this rate by the total live time gives the expected number of accidental coincidences, which is compared to the number of measured coincidences in Figure \ref{fig:barchart1}.  This figure shows a statistically significant correlation between events in the muon detector and the bolometers.

The usual CUORICINO \bbd ~analysis includes an anti-coincidence cut which excludes any bolometer event that occurs within 100 ms of any other bolometer event.  The bolometer anti-coincidence condition is used to reduce background, since the \bbd ~signal is expected to appear only in one bolometer.  Limiting the analysis to single-bolometer events, the number of coincidences between the muon and bolometer events is consistent with the number of expected accidentals, as shown in Figure \ref{fig:barchart2}.  Evidently, the bolometer anti-coincidence cut is very effective at eliminating potential muon-induced backgrounds.

The numbers of expected accidental and measured coincidences shown in Figure~\ref{fig:barchart2} provide an upper limit on the muon-induced contribution to the CUORICINO background.  These results are summarized in Table~\ref{tab:upperlimits}.  
The limits were computed by using the Feldman-Cousins method \cite{feldman-98} to obtain an upper limit, $\nu^{\rm up}$, on the expected number of muon-correlated signal events.  This number was converted into an upper limit on the background rate, $R^{\rm up}$, in the usual units of \bg ~as follows:
\begin{equation}
R^{\rm up} = \nu^{\rm up} \cdot \frac{1}{f_{\rm obs}} \cdot \frac{1}{X} \cdot 
\frac{1}{\Delta E}
\end{equation}
Here, $X = 3.99$ kg$\cdot$yr is the total exposure (active bolometer mass times live time) and $\Delta E$ is the size of the energy window.  The error on the energy window is taken to be on the order of the energy resolution, 7--9 keV on average.  The factor $f_{\rm obs} = 13.6 \pm 1.6 \%$ is the fraction of the muons producing signal in bolometers that are also observed in the scintillators.  It is obtained from the simulation described in Section \ref{sec:simulation} by taking the number of muon events which hit the bolometers {\it and} the scintillators divided by the total number of generated muon events which hit the bolometers.  The uncertainty in $f_{\rm obs}$ is the dominant systematic uncertainty in the conversion from $\nu^{\rm up}$ to $R^{\rm up}$; however, this uncertainty is much smaller than the statistical uncertainty, and has therefore been neglected in computing upper limits.  After applying the bolometer anti-coincidence cut, the upper limit on the muon-induced contribution to the CUORICINO background in the \bbd ~region of interest is 0.0021~\bg ~at 95\% confidence level.
 
\begin{table}
\begin{center}
\begin{tabular}{ c c c c c c }
\toprule
\multirow{2}{*}{Energy} & 
\multirow{2}{*}{$\langle A \rangle$} & 
\multirow{2}{*}{M} &
\multirow{2}{*}{~} &  \multicolumn{2}{c} {Upper Limits (95\% CL)}\\ \cmidrule(lr){5-6}
& & & & $\nu^{\rm up}$ & $R^{\rm up}/10^{-3}$ \\
\midrule
Low (200-400 keV) & 15.3 & 15 & ~ & 9.0  & 83 \\
Mid (400-2000 keV) & 16.7 & 17 & ~ & 10. & 12 \\
High (2-4 MeV) & 1.01 & 0 & ~ & 2.3 & 2.1 \\ 
\bottomrule
\end{tabular}
\caption{Upper Limits (95\% CL) on the contribution of muon-induced events to the CUORICINO background.  Limits were computed using the Feldman-Cousins method.  $\langle A \rangle$ is the expectation value of the number of accidental coincidences.  $M$ is the number of measured coincidences.  $\nu^{\rm up}$ is a limit on the mean number of observed muon-correlated signal events, while $R^{\rm up}$ gives an upper limit on the rate in \bg.}
\label{tab:upperlimits}
\end{center}
\end{table}

In principle, a muon (or spallation neutron) could produce long-lived ($T_{1/2} \gtrsim$ 50 ms) radioactive isotopes which could then decay producing a delayed coincidence signal.  Based on the small number of muon events and large background, we do not expect to be sensitive to this effect.  Consistent with this expectation, we find no evidence of a delayed coincidence signal.  However, due to the poor sensitivity and large number of potential products (each with a different half-life and decay energy), we do not set an upper limit for delayed coincidences with the present data.

\subsection{Simulation Results}
\label{sec:sim_results}

The analysis of the simulated events was carried out in the same way as for the actual measurements. The spectrum of muon events in the various scintillators appears to be correctly reproduced in simulations. The spectrum of bolometer events was divided into the same three energy regions: 200--400 keV, 400--2000 keV, and 2000--4000 keV.

\begin{table}
\begin{center}
\begin{tabular}{ c c c c }
\toprule
 & Simulation                     & Measurement \\
       & $10^{-3}$ (\bg)    & $10^{-3}$ (\bg)\\
 \multicolumn{3}{l} {All Events} \\
\midrule
Low (200-400 keV)  &  $25.0\pm0.7$ & $10\pm7$\\
Mid (400-2000 keV) &  $7.91\pm0.14$  & $4.2\pm1.1$\\
High (2-4 MeV)     &  $1.71\pm0.12$  & $1.2\pm0.4$\\
 & & \\
\multicolumn{3}{l} {With Bolometer Anti-coincidence Cut} \\
\midrule
Low (200-400 keV)  & $1.84\pm0.19$ & $< 11$ \\
Mid (400-2000 keV) & $0.66\pm0.04$ & $< 1.6$ \\
High (2-4 MeV)     & $0.08\pm0.03$ & $< 0.29$ \\
\bottomrule
\end{tabular}
\caption{Simulated and measured rates of bolometer events in coincidence with the muon detector. Only statistical errors are quoted. Systematic uncertainties are discussed in the text (Sections \ref{sec:simulation} and \ref{sec:dat_ana}).}
\label{tab:SimResVeto}
\end{center}
\end{table}

\begin{table}
\begin{center}
\begin{tabular}{ c c c c }
\toprule
Energy & Total                     & Anti-coincidence \\
       & $10^{-3}$ (\bg)    & $10^{-3}$ (\bg)\\
\midrule
Low (200-400 keV)  &  $184.9\pm1.9$ & $7.9\pm0.4$\\
Mid (400-2000 keV) &  $58.1\pm0.4$  & $3.58\pm0.09$\\
High (2-4 MeV)     &  $12.6\pm0.3$  & $0.53\pm0.06$\\
\bottomrule
\end{tabular}
\caption{Simulated contribution of muon-induced events to the CUORICINO background. Only statistical errors are quoted. Systematic uncertainties are discussed in the text (Section \ref{sec:simulation}).}
\label{tab:SimRes}
\end{center}
\end{table}

In Table \ref{tab:SimResVeto}, the simulated rates of bolometer events in coincidence with the muon detector are reported and compared with data (with and without imposing a bolometer anti-coincidence cut).  The measured rates are reported after the subtraction of the expected background from accidental coincidences.  In Table \ref{tab:SimRes}, the simulation results are reported for the total muon-induced background rate in CUORICINO.
In the energy region immediately surrounding the \bbd ~Q-value (2507.5--2547.5 keV), a value of  (17.4 $\pm$ 1.3)$\times$10$^{-3}$ \bg ~was obtained for background induced by muons without any anti-coincidence cut applied and a value of (0.61 $\pm$ 0.25)$\times$10$^{-3}$ \bg ~with the bolometer anti-coincidence cut.

\section{Conclusions}
The bolometer anti-coincidence cut in CUORICINO appears to be a very effective tool for eliminating muon-induced backgrounds.  With this cut, the measured rate of muon-correlated, single-bolometer background events was consistent with zero, and an upper limit of 0.0021 \bg ~(95\% C.L.) in the \bbd ~region of interest was obtained.

The results of the measurement have been compared with a detailed \textsc{Geant4} simulation. Although the sensitivity of the experiment was not sufficient to perform a rigorous validation, the results of the measurement and simulation were generally compatible.  

The rate obtained for the muon-induced contribution to the CUORICINO background, by measurement or simulation, is small compared to the total CUORICINO background rate of $\sim$0.2 \bg ~in the region of interest.  Muon interactions also do not appear to contribute significantly to the background rate between 3--4 MeV.  

The muon-induced backgrounds may not scale directly from CUORICINO to CUORE because of differences in the detector and shield geometry, materials, and anti-coincidence efficiency. For that reason, a detailed simulation, similar to that described in Section \ref{sec:simulation}, has also been performed for muons and other external backgrounds in CUORE~\cite{SimulationCUORE}.  However, omitting subtle changes, the muon-induced background rates in CUORE should be of a similar order of magnitude to those obtained for CUORICINO.  The CUORE goal for the total background rate in the region of interest is 0.01 \bg, and both the measured and simulated values for the muon-induced background in CUORICINO are well below the CUORE goal.

\section{Acknowledgments}
This work was supported by the US Department of Energy under contract numbers DE-AC52-07NA27344 at LLNL and DE-AC02-05CH11231 at LBNL, and by the INFN of Italy.  We also wish to thank Dr.~Joel Rynes of the US Department of Homeland Security, Jos\'{e} Angel Villar of the Universidad de Zaragoza, and Pierre Lecomte from Eidgenossische Tech. Hochschule Z\"{u}rich (ETHZ), Switzerland for the loan of plastic scintillator detectors used in the measurements reported here.

\end{linenumbers}

\end{document}